\documentclass{ifacconf}

\usepackage{graphicx}      %
\usepackage{natbib}        %
\usepackage{tabularx}     %

\usepackage{booktabs}      %
\usepackage{multirow}      %
\usepackage{siunitx}       %
\usepackage{subcaption}
\usepackage{amsmath}
\usepackage{amssymb}
\begin{document}

\onecolumn
\thispagestyle{empty}

\vspace*{1cm}
{\Huge \noindent IFAC Copyright Notice \par}
\vspace{1.2cm}

\noindent
\setlength{\fboxsep}{18pt}%
\setlength{\fboxrule}{0.6pt}%
\fbox{%
  \begin{minipage}{\dimexpr\linewidth-2\fboxsep-2\fboxrule\relax}%
    \large
    \copyright~2026 the authors. This work has been accepted to IFAC for publication under a Creative Commons Licence CC-BY-NC-ND.\\[2em]

    Accepted to be published in: IFAC-PapersOnLine, Proceedings of the 24th IFAC World Congress, Busan, Republic of Korea, August 2026.\\[2em]

    DOI: to be added upon publication.%
  \end{minipage}%
}%

\vfill
\clearpage
\twocolumn

\begin{frontmatter}

\title{Time-Series Anomaly Detection for Mobile Robots in Automotive Active Safety Testing using an RNN-VAE} 

\author[First,Second]{Henrik Meyer}  
\author[Second]{Karsten Raguse}
\author[Third]{Armando Walter Colombo}
\author[First]{Thomas Seel}
\author[First]{Simon F. G. Ehlers}

\address[First]{Leibniz University Hannover, Institute of Mechatronic Systems, An der Universität 1, 30823 Garbsen, Germany}
\address[Second]{Volkswagen AG, Berliner Ring 2, 38436 Wolfsburg, Germany (e-mail: henrik.meyer1@volkswagen.de).}
\address[Third]{University of Applied Sciences Emden/Leer, Constantiaplatz 4, 26723 Emden, Germany}

\begin{abstract}                %
Mobile robots, like the ultra-flat overrunable (UFO) robot platform, used in automotive active safety tests, currently lack self-diagnostic capabilities necessary to detect present hardware defects. This circumstance can lead to more severe failures, causing expensive repairs and operational downtime.
This work proposes, for the first time, a reconstruction-based time-series anomaly detection model for these mobile robots, considering defect classes such as unevenly worn full-rubber tires or damaged dampers. Unlike prior publications, the proposed approach leverages the vast quantities of unlabeled data generated during routine operation through a simple pre-training step. Furthermore, it optimizes the hyperparameters of the implemented gated recurrent unit-based variational autoencoder (GRU-VAE) and evaluates both a stateless, windowed training approach and one using truncated backpropagation through time (TBPTT).
The model’s generalization capabilities are demonstrated by successfully detecting six defect types, with four of them not present in the data used for hyperparameter optimization and threshold selection. This is validated using a test set collected from five system instances at various points over a period of several months, achieving an F1 score of \SI{0.936}{}, indicating strong practical viability.

\end{abstract}


\end{frontmatter}

\section{Introduction}

During a car's development and its subsequent evaluation by authorities, various practice tests are conducted to validate active safety systems, such as the emergency brake assist. In these tests, so-called target carriers, such as the ultra-flat overrunable (UFO) robot platform, are used to mimic other road users, like cars. These systems are subject to high strain while only being inspected irregularly, requiring manual intervention. Undetected initial defects, such as unevenly worn tires, can lead to more severe failures, resulting in expensive repairs and significant operational downtime of the equipment.

The identification of defects to increase the reliability of technical systems is one of the reasons why condition monitoring and anomaly detection have long been a subject of significant interest. One of the domains within this field is mobile robots. In this domain, work has mainly been done on the condition monitoring of individual components. One example is the work of \cite{Yu.2012}, where defects in the steering system of a mobile robot are identified using a model-based approach.
The often-present scarcity of labeled data, however, makes it impractical to implement dedicated condition monitoring models for every possible defect type. Because of this, the focus is shifted towards anomaly detection, where any deviations from normal behavior are detected without the need for prior knowledge of each specific defect.

Model-based anomaly detection methods, like the one from \cite{Guo.2018}, rely on the correct representation of system dynamics to detect misbehavior. While these approaches can be effective, they are often impractical to implement for complex systems. Supervised learning approaches, like the one from \cite{Zhang.2024}, where defects are injected into real data, rely on the prior knowledge of defect characteristics. This limits their ability to detect novel anomalies. Because of this, data-driven, unsupervised, or semi-supervised approaches have shifted into focus. While this field of time-series anomaly detection is broad, with performance being highly application-dependent (\cite{ZamanzadehDarban.2025}), reconstruction-based models like recurrent neural network (RNN)-based autoencoders (AEs) have proven suitable in applications related to the present use case.
One notable example is the work of \cite{Park.2018}, where deviations in the behavior of a feeding robot are detected based on the reconstruction error of a long short-term memory (LSTM)-based variational autoencoder (VAE). Another example is the work of \cite{Alizadeh.2023}, where a similar LSTM-AE-based approach is used to detect anomalies within different subsystems of a vehicle. Like other reconstruction-based approaches, they offer feature-level interpretability, offering insights for root cause analysis, while often being computationally efficient enough for inference on end-devices.
While the general approaches have proven suitable, practical challenges like (i) high variability within normal system behavior and (ii) a lack of normal (non-faulty) data are not addressed. Furthermore, (iii) hyperparameter optimization (HPO) of the proposed models is rarely considered, and (iv) validation is frequently performed with simulation data, or under idealized laboratory conditions, not taking into account domain shifts, like transfer to a different system instance or changing environmental conditions.

In this context, the objective of this paper is to equip the mobile robots used in automotive active safety tests (UFO) with the self-diagnostic capabilities necessary to detect present defects. To address the identified limitations within this application, the contributions of this work are: (i) the implementation of a gated recurrent unit (GRU)-based VAE coupled with a K-nearest-neighbor (KNN) to account for the variability of normal system behavior, (ii) the assessment of a simple two-stage training approach (pre-training on unlabeled data) to mitigate the scarcity of labeled, normal data, (iii) HPO of the models, including the comparison of a stateless and a stateful approach using truncated backpropagation through time (TBPTT), and (iv) the validation of the models using real-world test data with novel defect types and on multiple system instances.

\section{Reconstruction-based Anomaly Detection} \label{ch: Reconstruction based Anomaly Detection}

\begin{figure}
\begin{center}
\includegraphics[width=8.4cm]{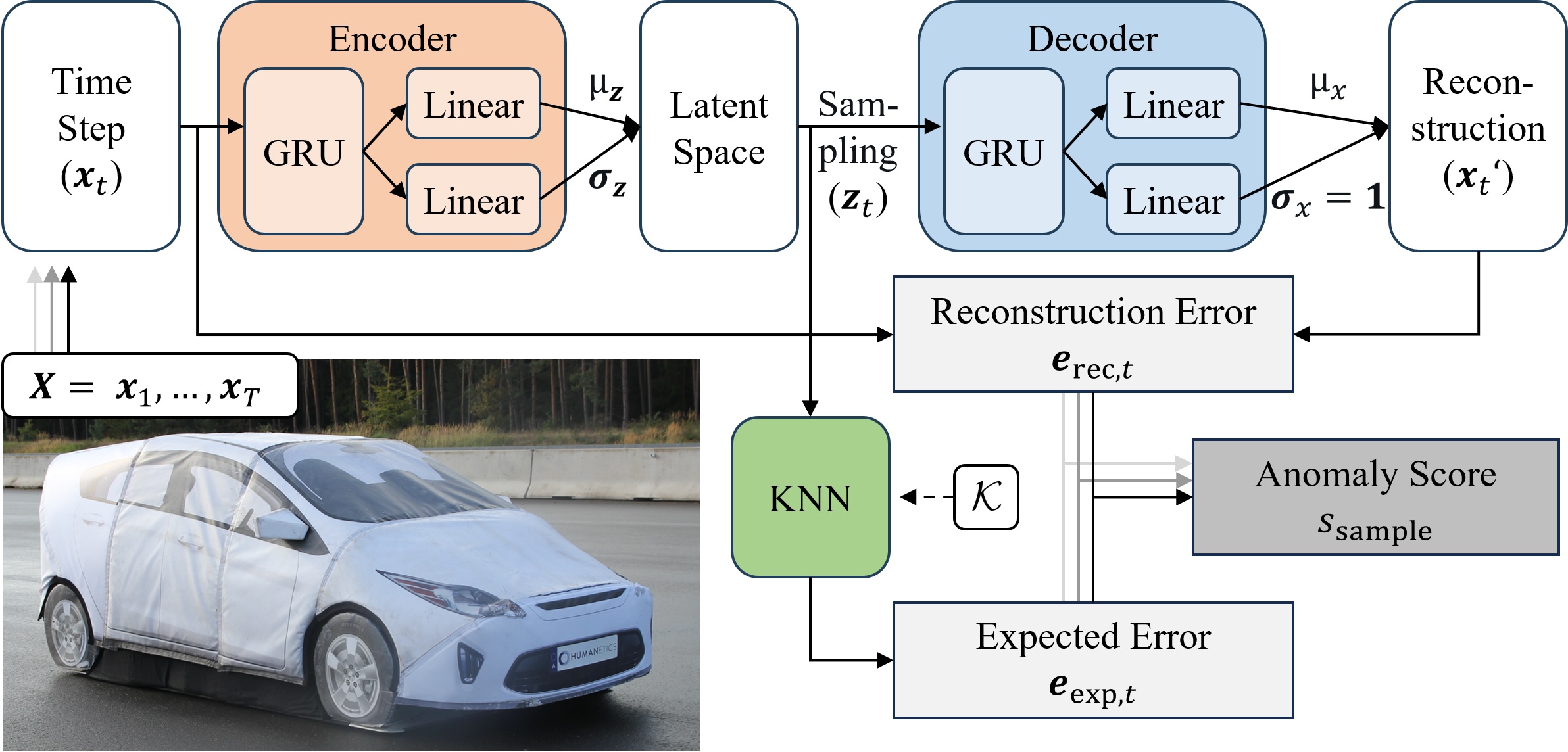}    %
\caption{Architecture of the proposed anomaly detection model, showing the processing of a time step $\boldsymbol{x}_t$ from a multivariate time-series $\boldsymbol{X}$. Elements denote model components (colored), input/output data (white), and calculated metrics (gray rectangles).}
\label{fig:Model}
\end{center}
\end{figure}

The core of the model is a GRU-VAE that reconstructs multivariate time-series. A KNN, trained on the latent space representations of normal samples, is used to predict the expected reconstruction error during inference. The final anomaly score is then calculated by taking the mean of the features with the highest ratio of reconstruction error to expected error. The proposed architecture can be seen in Fig. \ref{fig:Model}.

\subsection{Recurrent Neural Network-based \\Variational Autoencoder (RNN-VAE)}

A VAE (\cite{Kingma.2013}) is a variant of an AE that transforms a high-dimensional input $\boldsymbol{x}$ into a lower-dimensional latent representation $\boldsymbol{z}$ using an encoder $q_{\boldsymbol{\phi}}(\boldsymbol{z}|\boldsymbol{x})$. A decoder $p_{\boldsymbol{\theta}}(\boldsymbol{x}|\boldsymbol{z})$ then uses this latent representation to reconstruct the original data as $\boldsymbol{x'}$.
In this case, the input, as $\boldsymbol{x}_t$, represents one time step from a full sample (a multivariate time-series) $\boldsymbol{X}$ with $F$ features.
A key difference from a standard AE is that the encoder does not directly output a vector $\boldsymbol{z}$ in the latent space. Instead, it outputs the parameters of a probability distribution, oftentimes, as in this paper, a normal distribution with a mean vector $\boldsymbol{\mu}$ and a standard deviation vector $\boldsymbol{\sigma}$. To enable backpropagation through this stochastic step, the latent vector $\boldsymbol{z}$ is calculated using a reparameterization trick. This is achieved by combining the learned parameters with a random vector $\boldsymbol{\epsilon}$, as
$\boldsymbol{z} = \boldsymbol{\mu} + \boldsymbol{\sigma} \odot \boldsymbol{\epsilon}$, where $\boldsymbol{\epsilon} \sim \mathcal{N}(\boldsymbol{0}, \boldsymbol{I})$, with $\boldsymbol{I}$ denoting the identity matrix.

The VAE optimizes the evidence lower bound (ELBO), which balances reconstruction accuracy and latent space regularization. The loss function, as defined by \cite{Kingma.2013}, thus consists of two components
\begin{equation} \label{eq:vae_loss_clean}
\begin{split}
\mathcal{L}_{\mathrm{VAE}}(\boldsymbol{\theta}, \boldsymbol{\phi}; \boldsymbol{x}) = & -\mathbb{E}_{q_{\boldsymbol{\phi}}(\boldsymbol{z}|\boldsymbol{x})}[\log p_{\boldsymbol{\theta}}(\boldsymbol{x}|\boldsymbol{z})] \\
& + D_{\mathrm{KL}}(q_{\boldsymbol{\phi}}(\boldsymbol{z}|\boldsymbol{x}) || p(\boldsymbol{z})),
\end{split}
\end{equation}
with $\boldsymbol{\phi}$ and $\boldsymbol{\theta}$ as the encoder and decoder parameters.

The first part of the loss function, $-\mathbb{E}_{q_{\boldsymbol{\phi}}(\boldsymbol{z}|\boldsymbol{x})}[\log p_{\boldsymbol{\theta}}(\boldsymbol{x}|\boldsymbol{z})]$, is the reconstruction loss. It forces the model to accurately reconstruct the input $\boldsymbol{x}$ by maximizing the expected log-likelihood $\log p_{\boldsymbol{\theta}}(\boldsymbol{x}|\boldsymbol{z})$. This expectation is computed by first sampling latent vectors $\boldsymbol{z}$ from the encoder's distribution $q_{\boldsymbol{\phi}}(\boldsymbol{z}|\boldsymbol{x})$ and then using the decoder to evaluate the probability of the original data $\boldsymbol{x}$ given $\boldsymbol{z}$. Minimizing this term (equivalent to maximizing the log-likelihood) forces the model to generate reconstructions that are as close as possible to the original input. For the sake of simplicity, and to ensure more stable training, this work assumes a fixed decoder standard deviation vector of $\boldsymbol{\sigma}_{\boldsymbol{x}} = \mathbf{1}$. The reconstruction loss thereby simplifies to the mean squared error (MSE) between the input $\boldsymbol{x}$ and the reconstruction $\boldsymbol{x'}$.
The second part, $D_{\mathrm{KL}}(q_{\boldsymbol{\phi}}(\boldsymbol{z}|\boldsymbol{x}) || p(\boldsymbol{z}))$, is the regularization part, in the form of the Kullback-Leibler (KL) divergence. This term acts as a constraint on the encoder, measuring how much the learned posterior distribution $q_{\boldsymbol{\phi}}(\boldsymbol{z}|\boldsymbol{x})$ deviates from a simple prior distribution $p(\boldsymbol{z})$. In this work, the prior is defined as a standard normal distribution $\mathcal{N}(\boldsymbol{0}, \boldsymbol{I})$. Minimizing the divergence promotes a smooth and continuous latent space, where similar inputs yield similar distributions.

As the name suggests, an RNN-VAE consists of the just-described VAE, but with RNNs replacing the standard feed-forward networks. This enables the model to capture temporal dependencies in the sequential data, as the RNNs maintain hidden states across time steps.
As traditional RNNs suffer from vanishing and exploding gradients, gated RNNs in the form of LSTMs, introduced by \cite{Hochreiter.1997}, or GRUs, introduced by \cite{Cho.2014}, are commonly used. In this work, a GRU is used because of the comparable performance but simpler structure compared to LSTMs (\cite{Chung.2014}).

\subsection{K-Nearest-Neighbor (KNN)}

The KNN algorithm (\cite{Cover.1967}) is utilized as a non-parametric regression model to estimate the expected reconstruction error $\boldsymbol{e}_{\mathrm{exp}}$ for any given input $\boldsymbol{x}_{t}$.
It is chosen due to its deterministic behavior, computational efficiency, and minimal need for hyperparameter tuning.
The "training" phase of the KNN consists of building a reference database by passing normal samples through the trained GRU-VAE. The reference dataset consists of $N$ samples, indexed by $i$. For each sample $\boldsymbol{x}_i$, the encoder's latent space representation $\boldsymbol{z}_i$ is recorded, and the corresponding reconstruction error $\boldsymbol{e}_{\mathrm{rec},i}$ is calculated as the squared error between the input $\boldsymbol{x}_i$ and the decoder's output $\boldsymbol{x}_i'$. The resulting pairs $(\boldsymbol{z}_i, \boldsymbol{e}_{\mathrm{rec},i})$, defining the set $\mathcal{K} = \{(\boldsymbol{z}_i, \boldsymbol{e}_{\mathrm{rec},i})\}_{i=1}^{N}$, are then stored in the database.

During inference, an input $\boldsymbol{x}_t$ is encoded into its latent vector $\boldsymbol{z}_t$. The KNN algorithm identifies the set $\mathcal{K}(\boldsymbol{z}_t)$, which is a subset of $\mathcal{K}$ that contains the $k$ latent vectors $\boldsymbol{z}_i$ that are closest to $\boldsymbol{z}_t$, based on their Euclidean distance. The expected reconstruction error is then calculated as
\begin{equation}
\label{eq:knn_regression} \boldsymbol{e}_{\mathrm{exp},t} = \frac{1}{k} \sum_{i=1}^{k} \boldsymbol{e}_{\mathrm{rec},i}.
\end{equation}

This expected error is intended to improve the robustness of the model, as it accounts for the GRU-VAE's non-uniform reconstruction performance by leveraging the regularized latent space, where similar maneuvers are expected to correspond to similar distributions.

\subsection{Calculation of Anomaly Score}

Since, in this work, entire samples are to be classified as anomalies, it is necessary to compute an overall anomaly score from the reconstruction error $\boldsymbol{e}_\mathrm{rec}$ calculated by the GRU-VAE and the expected error $\boldsymbol{e}_\mathrm{exp}$ estimated by the KNN, as shown in Fig.~\ref{fig:Model}.

First, for each time step $t$ and feature $f$ of the considered sample $\boldsymbol{X}$, an error ratio is calculated as
\begin{equation}
r_{t,f} = \frac{e_{\mathrm{rec},t,f}}{e_{\mathrm{exp},t,f}}.
\end{equation}
Then, for each feature, a feature-specific anomaly score is calculated as
\begin{equation}
s_f = \frac{1}{|T_f^{\mathrm{top}}|} \sum_{t \in T_f^{\mathrm{top}}} r_{t,f},
\end{equation}
where $T_f^{\mathrm{top}}$ is the set of top \SI{50}{\%} time steps with the highest error ratio $r_{t,f}$ for the feature $f$.
Finally, the anomaly score of the overall sample $\boldsymbol{X}$ is calculated as
\begin{equation}
s_{\mathrm{sample}} = \frac{1}{|F^{\mathrm{top}}|} \sum_{f \in F^{\mathrm{top}}} s_f,
\end{equation}
where $F^{\mathrm{top}}$ is the set of the three features with the highest feature-specific anomaly scores.

The specific values for the ratio of time steps, as \SI{50}{\percent}, and the number of features, as three, are based on domain knowledge regarding typical defect characteristics (e.g., how many features a defect usually affects) and confirmed during initial, non-systematic tests with validation data. While these values can be optimized as hyperparameters, they are treated as constants in this work.

\section{Experimental Setup}

\begin{figure}
\begin{center}
\includegraphics[width=8.4cm]{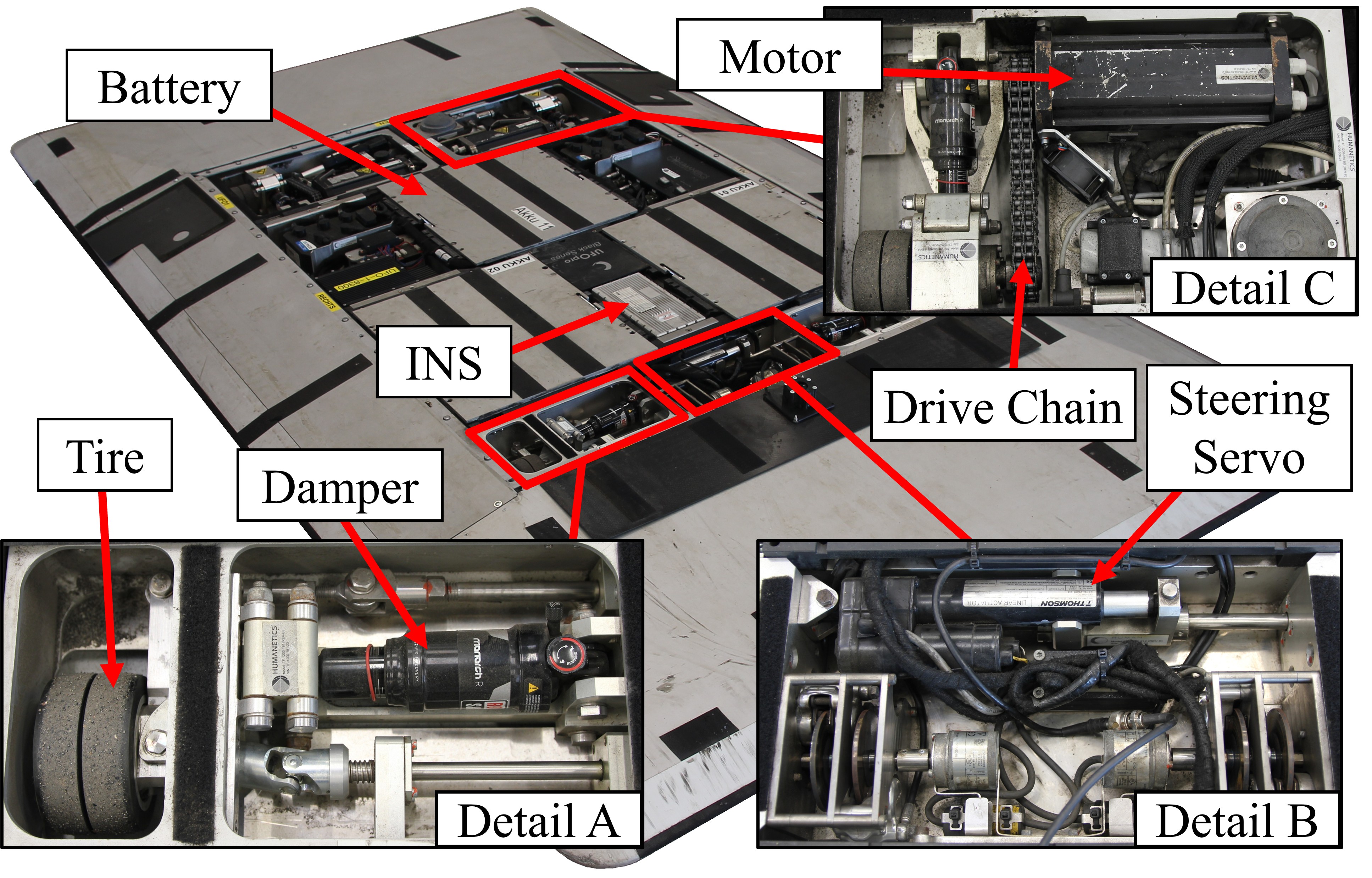}
\caption{UFO without cover plate.} 
\label{fig:UFOpro}
\end{center}
\end{figure}

The UFO robot platform (see Fig. \ref{fig:UFOpro}) consists of a core unit and ramps attached to each side, which enable cars, in the case of crashes, to drive over the platform without significant damage. With ramps attached, the mobile robot platform is \SI{2.95}{m} long, \SI{1.69}{m} wide, \SI{0.098}{m} high, and weighs \SI{264}{kg}. It is equipped with a standardized target, in this case a soft-car, that weighs an additional \SI{110}{kg}. The test-ready setup can be seen in Fig.~\ref{fig:Model}.

\begin{figure}
\begin{center}
\includegraphics[width=8.4cm]{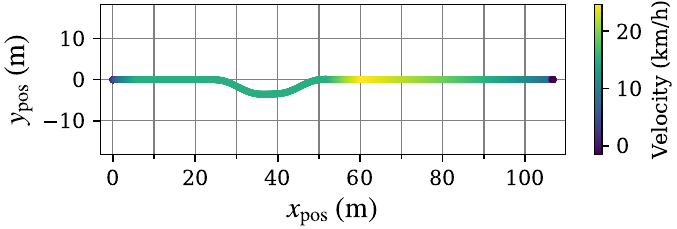}
\caption{Example of position and velocity within a checkup-run.} 
\label{fig:Checkup run}
\end{center}
\end{figure}

During daily operation, the UFO is used in different test procedures, following predefined trajectories with velocities of up to \SI{100}{km/h}. Position and velocity following one of these trajectories, the so-called checkup-run, can be seen in Fig. \ref{fig:Checkup run}. The checkup-run takes around \SI{25}{s} and is performed on a regular basis, for example, after any maintenance, to confirm that the system is in proper condition. The data recorded following this trajectory forms the data basis for this work. Parts of this dataset originate from tests where defects are manually induced, while other parts come from the regular tests, where the condition of the system is known. To ensure diversity in the dataset, tests are conducted with five different UFO instances, at various positions on the test site, and at different times of the year, under changing weather conditions.

In each predefined test run, data is recorded at a rate of $\SI{100}{Hz}$. In this work, a subset of $F = 34$ features is used. Their scope is based on domain knowledge and includes data from the front and rear wheel speed sensors; velocity and acceleration from the GPS module; brake parameters, such as brake pressure and brake servo position; battery parameters, such as voltage and current; and steering parameters, such as steering angle and steering servo position. In addition to these sensor values, the dataset also contains control values, such as the desired velocity, and values derived from the sensors, such as tire slip.

\section{Model Training and Hyperparameter Optimization}

Before training and optimizing the models, the datasets used for the different training steps are introduced. All of the data, except the pre-training set, is recorded following the checkup-run trajectory.

\begin{itemize}
    \item \textbf{Pre-training set:} Consists of \SI{250}{} unlabeled samples $\boldsymbol{X}$, randomly drawn from \SI{2,000}{} test runs. This set includes various trajectories from all six UFO instances, with the system's condition being unknown. The subset is used here to reduce computation time.
    
    \item \textbf{Training set:} Contains \SI{66}{} labeled normal (non-faulty) samples, recorded from UFO instances \SI{2}{} and \SI{4}{} immediately following extensive annual maintenance, representing an ideal condition.

    \item \textbf{KNN set:} Contains \SI{20}{} labeled normal samples from UFO instances \SI{1}{} and \SI{3}{}, recorded during normal operations, representing a good, but less-than-perfect, state.

    \item \textbf{Validation set:} A labeled set of \SI{30}{} samples (\SI{15}{} normal, \SI{15}{} anomalous), recorded from UFO instances 1, 2, and 3. This set contains two different defect types, steering (play in steering) and tire (unevenly worn tires), and is used for hyperparameter optimization and threshold selection.

    \item \textbf{Test set:} A labeled set of \SI{100}{} samples (\SI{50}{} normal, \SI{50}{} anomalous), from five UFO instances (1-5), used for model evaluation. It includes the two defect types from the validation set, plus four additional, novel ones: damper (insufficient pressure), rear wheel speed sensor (sensor drift), front wheel speed sensor (loose mounting), and battery (short circuit in one cell).
\end{itemize}

The anomaly detection model is trained in several configurations, based on three primary distinctions.
First, the models differ in how the GRU's hidden state is managed. In the stateless approach, the data is split into fixed-length, overlapping windows (hop size of one). The GRU's hidden states are only passed within these windows and reset afterwards. While the model is trained to reconstruct the entire window, only the reconstruction of the last time step is utilized for evaluation during inference.
Conversely, in the stateful approach using TBPTT, the time-series are split into non-overlapping windows (where the hop size equals the window size). Here, the GRU's hidden state is passed from one window to the next, and is only reset after each full sample.
The second point of differentiation is the data used to fit the scaler. The standard scaler that is used for preprocessing is fit either on the train data or the pre-train data.
The third distinction is the training strategy. While all approaches optimize the same self-supervised reconstruction loss (Eq.~\ref{eq:vae_loss_clean}), they utilize different datasets. "Standard training" uses only the training data, "pre-training" uses exclusively the unlabeled pre-training data, and "fine-tuning" takes a pre-trained model and continues its training on the training data.
A naming convention based on these three distinctions is presented in Table \ref{tab:model_naming}.

\begin{table}[htbp]
\centering
\caption{Model naming convention.}
\label{tab:model_naming}
\begin{tabular}{@{} l l l @{}}
\toprule
\textbf{Category} & \textbf{Code} & \textbf{Description} \\
\midrule
State management & stl & Stateless \\
                 & stf & Stateful (TBPTT) \\
\addlinespace
Scaler      & ts  & Fitted on training data \\
                 & ps  & Fitted on pre-train data \\
\addlinespace
Training strategy & std & Trained on train data \\
                  & pt  & Pre-trained only \\
                  & ft  & Pre-trained + fine-tuned \\
\bottomrule
\multicolumn{3}{@{}l@{}}{
    Example: \texttt{stl\_ps\_ft}. } \\
\end{tabular}
\end{table}

The models are optimized using a tree-structured Parzen estimator (TPE) (\cite{Bergstra.2011}). While the goal in each trial is to lower the VAE's cost function, the overall optimization objective is to maximize the area under the receiver operating characteristic curve (AUROC). It is calculated at the end of each trial on the validation set, using the full model described in Section \ref{ch: Reconstruction based Anomaly Detection}.

Four configurations are selected for HPO, two stateless models (stl\_ts\_std, stl\_ps\_std) and two stateful models using TBPTT (stf\_ts\_std, stf\_ps\_std). The hyperparameters found for the standard configurations are later used to train their corresponding pre-trained and fine-tuned counterparts. This is done to reduce computation time. In each case, the parameters shown in Table \ref{tab:hpo_parameters} are optimized.

\begin{table}[htbp]
\centering
\caption{Hyperparameters for optimization.}
\label{tab:hpo_parameters}
\begin{tabular}{@{} l l @{}}
\toprule
\textbf{Hyperparameter} & \textbf{Value range} \\
\midrule

Window size & 20, 50, 100 \\
Hidden size & 128, 256, 384, 512 \\
KL divergence weight & 0.01 ... 1.0 \\
Learning rate & \num{1e-5} ... \num{1e-3} \\
\bottomrule

\end{tabular}
\end{table}

The value ranges for window size and hidden size are limited to reduce the training time. Other parameters are excluded from the HPO after initial trials have shown that they have a comparatively small effect on classification performance. The batch size is set to \SI{128}{}, the latent space dimension to \SI{3}{}, and an L2 regularization is set to zero.
Each optimization is conducted for \SI{50}{} trials. Within each trial, the model is trained for a maximum of \SI{50}{} epochs, utilizing early stopping with a patience of \SI{5}{} and a manually set minimum delta to prevent overfitting on the small training set. As an example, the optimized parameters for the stl\_ps\_std model are a window size of \SI{100}{}, a hidden size of \SI{512}{}, a KL divergence weight of \SI{0.372}{}, and a learning rate of \SI{2.508e-5}{}.

\section{Results of Anomaly Detection}

\begin{figure}[htbp]
\centering

\begin{subfigure}{\columnwidth}
    \centering
    \includegraphics[width=8.4cm]{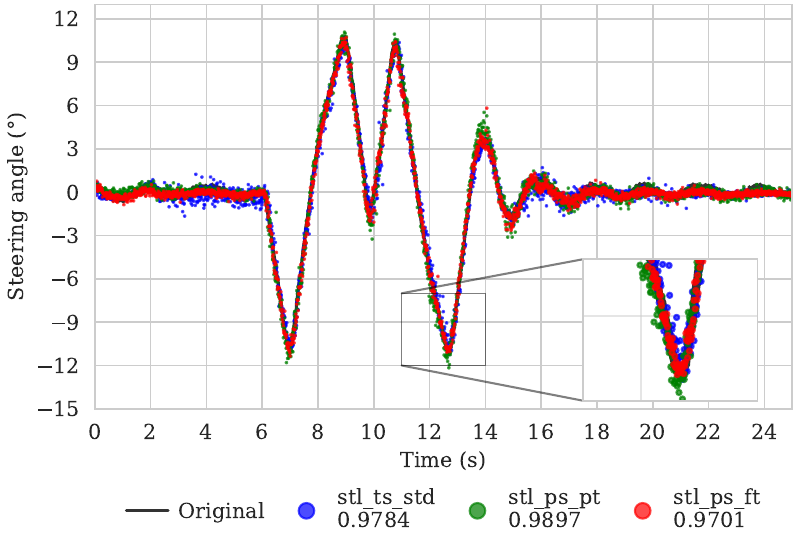}
    \caption{Normal sample from training set}
    \label{fig:subfigure_a}
\end{subfigure}
\\ [1ex]

\begin{subfigure}{\columnwidth}
    \centering
    \includegraphics[width=8.4cm]{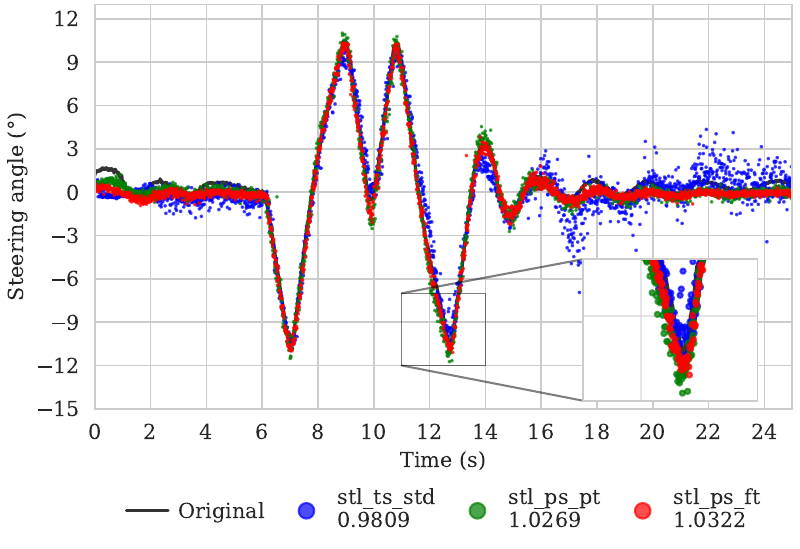}
    \caption{Normal sample from test set}
    \label{fig:subfigure_b}
\end{subfigure}
\\ [1ex] 

\begin{subfigure}{\columnwidth}
    \centering
    \includegraphics[width=8.4cm]{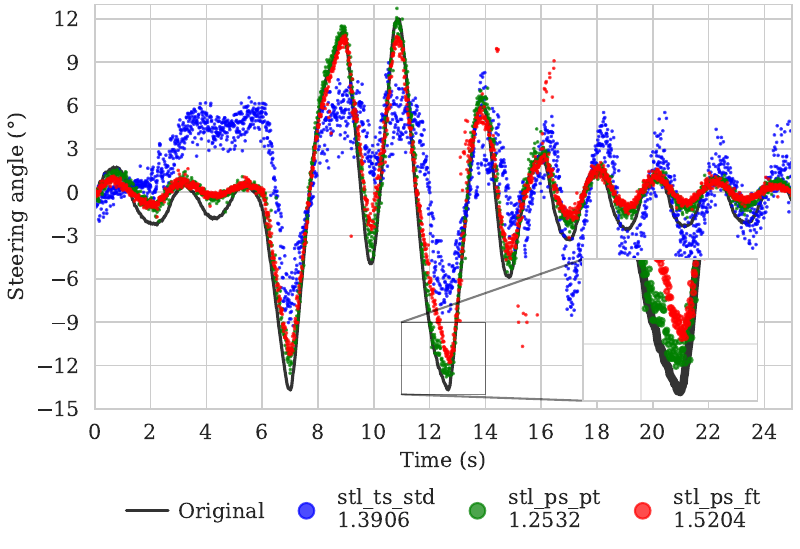}
    \caption{Anomalous sample with steering defect from test set}
    \label{fig:subfigure_c}
\end{subfigure}

\caption{Reconstruction of steering angle with feature-specific anomaly scores $s_f$.}
\label{fig:Reconstruction example}
\end{figure}

To exemplify the model's behavior, the reconstructions of the steering angle are examined for three different samples $\boldsymbol{X}$, as shown in Fig. \ref{fig:Reconstruction example}. Although only the stateless models are assessed here, similar findings can be obtained for their stateful counterparts.
While all models are able to reconstruct normal data from the training set reasonably well, as shown in Fig. \ref{fig:subfigure_a}, differences can be seen when looking at a test set example. Here, the standard model (stl\_ts\_std) already deviates significantly from the actual sensor signal, even though the test sample, shown in Fig. \ref{fig:subfigure_b}, stems from a system without defects.
When taking a look at an anomalous test sample, as shown in Fig. \ref{fig:subfigure_c}, it can be seen that the reconstruction from the standard model deviates significantly from the actual signal. Furthermore, it is evident that the pre-trained model (stl\_ps\_pt) reconstructs the steering angle very accurately in most sections. The fine-tuned model, in contrast, deviates further and is closer to a signal as it would look without the present defect.

This observable difference between the shown models also translates to the feature-specific anomaly scores $s_f$. When calculating the score difference between the normal test sample and the anomalous one, the stl\_ts\_std model shows a ratio of $1.3906 / 0.9809 \approx 1.418$ (see legends in Fig. \ref{fig:Reconstruction example} b/c), which translates into a $\SI{41.8}{\percent}$ increase, while the stl\_ps\_pt model shows an increase of $\SI{22.0}{\percent}$, and the stl\_ps\_ft model one of $\SI{47.3}{\percent}$. In this context, a higher increase between normal and anomalous samples can be considered better. This is because the goal is to reconstruct normal data well, but not to reconstruct anomalies, so that differences between anomalous and normal behavior become apparent through the reconstruction error.
It should be noted that the values presented are derived from the overall models. This explains the relatively low anomaly scores of the standard model (stl\_ts\_std). By using the expected error $\boldsymbol{e}_{exp}$ from the KNN to calculate the feature-specific anomaly score $s_f$, the standard model already “knows” that it cannot reconstruct this feature particularly well, even when looking at normal samples.

An overall performance comparison for the optimized models can be seen in Table \ref{tab:model_comparison_simple}.
The fine-tuned, stateless model (stl\_ps\_ft) achieves the best performance on both the validation and the test sets. The stateful models perform slightly worse on average, despite their theoretical advantage of having a larger context window, and their observable advantage of producing few to no outliers when looking at their reconstructions.

\begin{table}[htbp]
\centering
\caption{Comparison of optimized models, with threshold for test set based on validation set, and model naming according to Table \ref{tab:model_naming}.}
\label{tab:model_comparison_simple}
\begin{tabular}{@{} l l l l l l @{}}
\toprule
\textbf{Model} & \textbf{Data} & \textbf{AUROC} & \textbf{F1} & \textbf{TPR} & \textbf{FPR} \\
\midrule

\multicolumn{6}{l}{\textbf{GRU-VAE: Stateless}} \\
\quad stl\_ts\_std & Val. & 0.791 & 0.750 & 0.600 & 0.000 \\
 & Test & 0.853 & 0.710 & 0.560 & 0.020 \\
\addlinespace

\quad stl\_ps\_std & Val. & 0.764 & 0.636 & 0.468 & 0.000 \\
 & Test & 0.944 & 0.734 & 0.580 & 0.000 \\
\addlinespace
\quad stl\_ps\_pt & Val. & 0.769 & 0.833 & 1.000 & 0.400 \\
 & Test & 0.902 & 0.763 & 0.900 & 0.460 \\
\addlinespace
\quad stl\_ps\_ft & Val. & \textbf{0.889} & \textbf{0.875} & 0.933 & 0.200 \\
 & Test & \textbf{0.961} & \textbf{0.936} & 0.880 & 0.000 \\
\addlinespace

\multicolumn{6}{l}{\textbf{GRU-VAE: Stateful (TBPTT)}} \\
\quad stf\_ts\_std & Val. & 0.884 & 0.839 & 0.867 & 0.200 \\
 & Test & 0.889 & 0.832 & 0.840 & 0.180 \\
\addlinespace

\quad stf\_ps\_std & Val. & 0.809 & 0.759 & 0.733 & 0.200 \\
 & Test & 0.910 & 0.739 & 0.680 & 0.160 \\
\addlinespace
\quad stf\_ps\_pt & Val. & 0.676 & 0.737 & 0.933 & 0.600 \\
 & Test & 0.930 & 0.838 & 0.880 & 0.220 \\
\addlinespace
\quad stf\_ps\_ft & Val. & 0.809 & 0.774 & 0.800 & 0.267 \\
 & Test & 0.953 & 0.857 & 0.780 & 0.040 \\

\bottomrule
\end{tabular}
\end{table}

\begin{figure}
\begin{center}
\includegraphics[width=8.4cm]{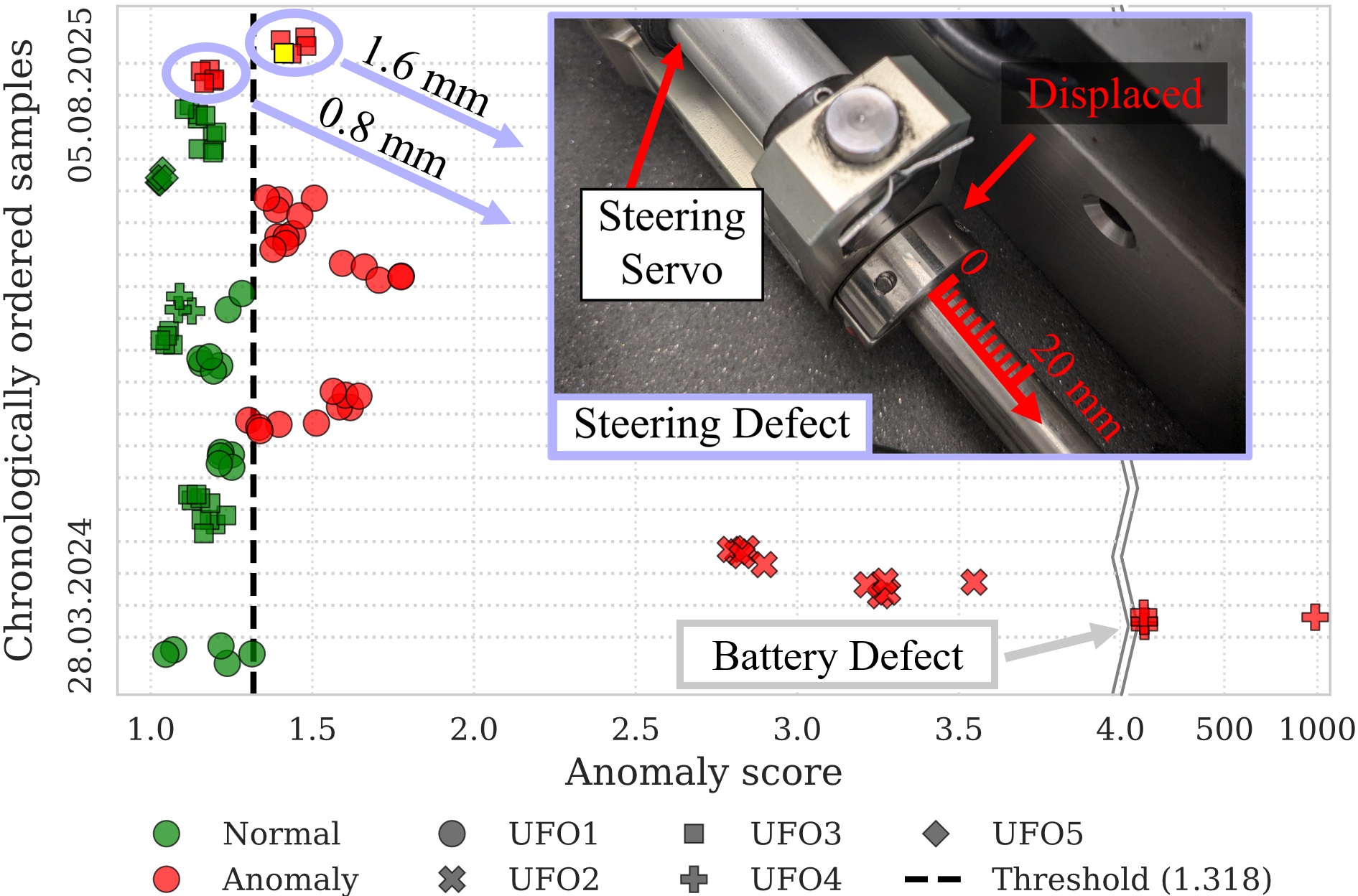}    %
\caption{Anomaly score distribution of test set with stl\_ps\_ft model. Grouped in batches of five samples recorded under similar conditions, steering defect from Fig. \ref{fig:subfigure_c} marked in yellow.} 
\label{fig:anomaly score distribution}
\end{center}
\end{figure}

The model's anomaly scores $s_{\mathrm{sample}}$ vary significantly, with some defects causing only a minor increase, while others result in scores far above the threshold.
It can be observed that tests conducted under similar conditions lead to similar anomaly scores. In Fig. \ref{fig:anomaly score distribution}, these test groups are visually separated by horizontal lines.
Furthermore, the magnitude of the anomaly score correlates with the defect's severity, considering the same defect type. This is demonstrated by the misclassified steering defect shown, which is correctly detected in a more severe version, where the play in steering was manually increased from \SI{0.8}{mm} to \SI{1.6}{mm} (see Fig. \ref{fig:anomaly score distribution}).

Another characteristic of the model becomes apparent when looking at the feature-specific anomaly scores $s_f$. For the majority of the anomalies examined, the top three features, used to calculate the score $s_{\mathrm{sample}}$, are directly related to the defect type. In the case of the unevenly worn tires, the most noticeable features always include vertical acceleration and the slip of the affected tire. In the case of the damper defect, the slip of the corresponding wheel has a high anomaly score.
On target hardware (Intel i5-1145G7 CPU, \SI{8}{GB} RAM), end-to-end inference for a \SI{25}{s} checkup-run takes approximately \SI{6}{s}, enabling on-device evaluation directly after testing.

As in most anomaly detection methods, the developed model is subject to limitations. One limitation is that (i) defects in the data must be clearly visible, i.e., they must exceed a deviation that could otherwise be caused by, for example, changing environmental conditions.
The sensitivity of the model (ii) also depends heavily on the model's ability to reconstruct various features well. If the model is unable to produce a good reconstruction, even for normal samples, differences caused by defects cannot be detected. 
Finally,  (iii) disturbances not caused by defects can also be classified as anomalies. In the presented model, normal tests conducted in heavy rain or without a target attached are classified as anomalies.

\section{Conclusion}

This work presents an anomaly detection model for mobile robots used in automotive active safety testing. The approach combines a GRU-VAE as a reconstruction-based model with a KNN to estimate an expected reconstruction error, enabling the calculation of a dynamic anomaly score.
The optimization of four models and comparison of eight model configurations reveal that a stateless approach outperforms the stateful alternative on average. Furthermore, the combination of pre-training on unlabeled data and fine-tuning on labeled, normal data significantly improves performance in this use case. Using this training strategy, an AUROC of \SI{0.961}{} is achieved, compared to a score of \SI{0.853}{} for the respective non-pre-trained version.
The model's robustness is demonstrated on a test set comprising \SI{100}{} labeled samples from five system instances, collected over several months, including four novel defect types.

In future work, the presented approach could be extended to make use of other trajectories, enabling the detection of anomalies within every possible test scenario. In addition, it should be compared against alternative methods that also make use of unlabeled data, such as generative adversarial networks or self-supervised contrastive approaches.

\begin{ack}
This research project is supported by Volkswagen AG. The results, opinions, and conclusions expressed in this publication are those of the authors and do not necessarily represent the views of Volkswagen AG.
\end{ack}

\bibliography{ifacconf}

@article{Alizadeh.2023,
 author = {Alizadeh, Morteza and Ma, Junfeng},
 year = {2023},
 title = {High-dimensional time series analysis and anomaly detection: A case study of vehicle behavior modeling and unhealthy state detection},
 volume = {57},
 number = {C},
 issn = {1474-0346},
 journal = {Adv. Eng. Inform.},
 doi = {10.1016/j.aei.2023.102041}
}

@article{Bergstra.2011,
 author = {Bergstra, James and Bardenet, R{\'e}mi and Bengio, Yoshua and K{\'e}gl, Bal{\'a}zs},
 year = {2011},
 title = {Algorithms for hyper-parameter optimization},
 volume = {24},
 journal = {Advances in neural information processing systems}
}

@misc{Cho.2014,
 author = {Cho, Kyunghyun and {van Merrienboer}, Bart and Bahdanau, Dzmitry and Bengio, Yoshua},
 year = {2014},
 title = {On the Properties of Neural Machine Translation: Encoder-Decoder Approaches},
 publisher = {arXiv},
 doi = {10.48550/ARXIV.1409.1259}
}

@misc{Chung.2014,
 author = {Chung, Junyoung and Gulcehre, Caglar and Cho, Kyunghyun and Bengio, Yoshua},
 year = {2014},
 title = {Empirical Evaluation of Gated Recurrent Neural Networks on Sequence Modeling},
 publisher = {arXiv},
 doi = {10.48550/ARXIV.1412.3555}
}

@article{Cover.1967,
 author = {Cover, Thomas and Hart, Peter},
 year = {1967},
 title = {Nearest neighbor pattern classification},
 pages = {21--27},
 volume = {13},
 number = {1},
 journal = {IEEE transactions on information theory}
}

@inproceedings{Guo.2018,
 author = {Guo, Pinyao and Kim, Hunmin and Virani, Nurali and Xu, Jun and Zhu, Minghui and Liu, Peng},
 title = {{RoboADS}: Anomaly Detection Against Sensor and Actuator Misbehaviors in Mobile Robots},
 pages = {574--585},
 booktitle = {2018 48th Annual IEEE/IFIP International Conference on Dependable Systems and Networks (DSN)},
 year = {2018},
 doi = {10.1109/DSN.2018.00065}
}

@article{Hochreiter.1997,
 author = {Hochreiter, Sepp and Schmidhuber, J{\"u}rgen},
 year = {1997},
 title = {Long short-term memory},
 pages = {1735--1780},
 volume = {9},
 number = {8},
 journal = {Neural computation}
}

@misc{Kingma.2013,
 author = {Kingma, Diederik P. and Welling, Max},
 year = {2013},
 title = {Auto-Encoding Variational {Bayes}},
 publisher = {arXiv},
 doi = {10.48550/ARXIV.1312.6114}
}

@misc{Park.2018,
 author = {Park, Daehyung and Hoshi, Yuuna and Kemp, Charles C.},
 year = {2018},
 title = {A Multimodal Anomaly Detector for Robot-Assisted Feeding Using an  {LSTM}-based Variational Autoencoder},
 url = {http://arxiv.org/pdf/1711.00614v1}
}

@inproceedings{Yu.2012,
 author = {Yu, Ming and Wang, Danwei and Chen, Qijun},
 title = {Prediction of multiple failures for a mobile robot steering system},
 pages = {1240--1245},
 booktitle = {2012 IEEE International Symposium on Industrial Electronics},
 year = {2012},
 doi = {10.1109/ISIE.2012.6237267}
}

@article{ZamanzadehDarban.2025,
 author = {{Zamanzadeh Darban}, Zahra and Webb, Geoffrey I. and Pan, Shirui and Aggarwal, Charu and Salehi, Mahsa},
 year = {2025},
 title = {Deep Learning for Time Series Anomaly Detection: A Survey},
 pages = {1--42},
 volume = {57},
 number = {1},
 issn = {0360-0300},
 journal = {ACM Computing Surveys},
 doi = {10.1145/3691338}
}

@article{Zhang.2024,
 author = {Zhang, Ze and Yao, Yue and Hutabarat, Windo and Farnsworth, Michael and Tiwari, Divya and Tiwari, Ashutosh},
 year = {2024},
 title = {Time Series Anomaly Detection in Vehicle Sensors Using Self-Attention Mechanisms},
 pages = {15964--15976},
 volume = {25},
 number = {11},
 journal = {IEEE Transactions on Intelligent Transportation Systems},
 doi = {10.1109/TITS.2024.3415435}
}

\end{document}